\documentclass{PoS}

\usepackage{amsmath}
\usepackage{amssymb}
\usepackage{bm}
\usepackage{cite}

\newcommand{\note}[1]{#1}

\title{Direct detection of  fourth generation Majorana neutrino dark matter }
\author{\speaker{Yu-Feng Zhou}\\ 
   State Key Laboratory of Theoretical Physics,\\
       Kavli Institute for Theoretical Physics China,\\
 Institute of  Theoretical Physics, Chinese Academy of Sciences,\\
  Beijing, 100190, P.R. China\\
  E-mail: \email{yfzhou@itp.ac.cn}
}

\ShortTitle{fourth generation neutrino dark matter} \abstract{ Heavy stable
  fourth generation Majorana neutrinos contribute to a small fraction of the
  relic density of dark matter (DM) in the Universe. Due to its relatively strong
  coupling to the standard model particles, it can be probed by the current
  direct and indirect DM detection experiments even it is a subdominant
  component of the  halo DM.  
  We show that the current Xenon100 data constrain the mass of the stable
  Majorana neutrino to be greater than the mass of the top
  quark. 
  The effective spin-independent cross section for the neutrino elastic
  scattering off nucleon is predicted to be $\sim 1.5\times 10^{-44} \mbox{
    cm}^2$, which is insensitive to the neutrino mass and mixing and can be
  reached by the direct DM detection experiments in the near future.  In the same mass
  region the predicted effective spin-dependent cross section for the heavy
  neutrino scattering off proton is in the range of $2\times 10^{-40} \mbox{
    cm}^2\sim 2\times 10^{-39}\mbox{ cm}^2$, which is within the reach of the
  ongoing \note{DM indirect search experiments.} We
  demonstrate such properties of the heavy neutrino DM in a fourth generation
  model with the stability of the fourth Majorana neutrino protected by an
 anomaly-free $U(1)$ gauge symmetry.  }

\FullConference{VIII International Workshop on the Dark Side of the Universe,\\
		June 10-15, 2012\\
		Rio de Janeiro, Brazil}

\begin{document}

\indent Models with chiral fourth generation fermions are among the simplest and
well-motivated extensions of the standard model (SM) and have been extensively
studied~\cite{Frampton:1999xi}.  The condition for CP symmetry violation in
the SM requires at least three generations of
fermions~\cite{Kobayashi:1973fv}. But, there is no upper limit on the
number of generations from the first principle.
In the SM the amount of CP violation is not large enough to explain the
baryon-antibaryon asymmetry in the Universe. The inclusion of fourth
generation quarks leads to two extra CP phases in quark sector and possible
larger CP violation~\cite{0803.1234}, which is helpful for electroweak baryogenesis.  
With very massive quarks in the fourth generation, it has been proposed that
the electroweak symmetry breaking may become a dynamical feature of the
SM~\cite{Holdom:1986rn,Carpenter:1989ij,Hill:1990ge,Hung:2009hy}.

Heavy stable neutrinos with mass greater than $\sim 1$ GeV are possible
candidates for the cold DM \cite{Lee:1977ua,Kolb:1985nn}.  However, if the
neutrino is the dominant component of the halo DM, the current DM direct
search experiments have imposed strong constraints on its mass~\cite{Goodman:1984dc,Srednicki:1986vj,Falk:1994es,0706.0526,Angle:2008we,Keung:2011zc}.
On the other hand, it is well-known that for a neutrino heavier than $ \sim
m_Z/2$, the cross section for its annihilation is in general too large to
reproduce the observed DM relic density. 
For the neutrino heavier than $m_W$, the contribution from $f\bar{f}$ channels
decrease rapidly, but other channels such as $W^\pm W^\mp$, $Z^0 h^0$
etc.  are opened. For these processes the corresponding cross section does not
decrease with the increasing of the neutrino mass, resulting in a relic
density always decreases with the growing of the neutrino mass, and a thermal
relic density far below the observed total DM relic
density~\cite{Enqvist:1988we}.
\note{ Thus the neutrino DM can only
  contribute to a small fraction of the relic density of DM and a small
  fraction of the halo DM density as well.
}
Despite its very low number density in the halo, it can still be probed
by the underground DM direct detection experiments due to its relatively strong coupling
to the target nuclei, which provides a way to search for new physics beyond
the SM complementary to the LHC. 

In this talk, we discuss the consequence of this possibility in a model with a
fourth generation Majorana neutrino DM. The stability of the fourth Majorana
neutrino protected by an additional generation-dependent $U(1)$ gauge
symmetry which is anomaly-free. The details of the analysis can be found in Ref. \cite{Zhou:2011fr}.
We consider a simple extension of the SM with a sequential fourth
generation and an additional $U(1)_F$ gauge symmetry. \note{The
  $U(1)$ extensions to the SM are well motivated from the point view of grand
  unification such as the $SO(10)$ and $E_6$ and have rich
  phenomenology~\cite{Langacker:2008yv} which can be reached by the on going  LHC
  experiments}. The flavor contents in the model are given by
\begin{align}   
q_{iL}=\begin{pmatrix} u_{iL} \\ d_{iL} \end{pmatrix}, \
\ell_{iL}=\begin{pmatrix} \nu_{iL} \\ e_{iL} \end{pmatrix}, \
u_{iR},\ d_{iR}, \ \nu_{iR}, \ e_{iR} \ (i=1,\dots, 4) .
\end{align}
All the fermions in the model are vector-like under the extra gauge
interactions associated with $U(1)_F$. The $U(1)_F$ charges of the fermions could  be  generation-dependent. In order to
evade the stringent constraints from the tree-level flavor changing neutral
currents (FCNCs), the $U(1)_F$ charges $Q_{qi}$ for the first three generation
quarks are set to be the same, i.e. $Q_{qi}=Q_q, \ (i=1,2,3)$ while $Q_{q4}=-3 Q_q$ for the
fourth generation quarks. Similarly, the $U(1)_F$ charges for the first three
generation and the fourth generation leptons are $Q_L$ and $-3 Q_L$,
respectively. In general, $Q_q$ and $Q_L$ can be different. For simplicity,  we consider $Q_q=Q_L=1$.
With this set of flavor contents and $U(1)_F$ charge assignments, it is
straight forward to see that the new gauge interactions are  anomaly-free. Since
the gauge interaction of $U(1)_F$ is vector-like, the triangle anomalies of
$[U(1)_F]^3$, $[SU(3)_C]^2 U(1)_F$ and $[\mbox{gravity}]^2 U(1)_F$ are all
vanishing.  The anomaly of $U(1)_Y [U(1)_F]^2$ is zero because the $U(1)_Y$
hypercharges cancel for quarks and leptons separately in each generation,
namely $\sum (- Y_{qL}+Y_{qR})=0$ and $\sum (- Y_{\ell L}+Y_{\ell R})=0$.  The
anomaly of $[SU(2)_L]^2 U(1)_F$ is also zero due to the relation
$\sum_{i=1}^{4}Q_{qi}=0 $ and $\sum_{i=1}^{4} Q_{Li}=0$. 
Thus in this model, the gauge anomalies generated by the first three
generation fermions are canceled by that of the fourth generation one, which
also gives a motivation for the inclusion  of the fourth generation.
 
The gauge symmetry $U(1)_F$ is to be spontaneously broken by the Higgs
mechanism. For this purpose we introduce two SM singlet scalar fields
$\phi_{a,b}$ which carry the $U(1)_F$ charges $Q_a=-2 Q_L$ and $Q_b=6 Q_L$
respectively. The $U(1)_F$ charges of $\phi_{a,b}$ are arranged such that
$\phi_a$ can have Majorana type of Yukawa couplings to the right-handed
neutrinos of the first three generations $\nu_{iR}\ (i=1,2,3)$ while $\phi_b$
only couples to the fourth generation neutrino $\nu_{4R}$. After the
spontaneous symmetry breaking, the two scalar fields obtain vacuum expectation
values (VEVs) $\langle \phi_{a,b}\rangle=v_{a,b}/\sqrt{2}$.
The relevant interactions in the model are given by
\begin{align}
\mathcal{L}&=
\bar{f}_i i\gamma^\mu D_\mu f_i 
+(D_\mu\phi_a)^\dagger (D_\mu\phi_a)+(D_\mu\phi_b)^\dagger (D_\mu\phi_b)
\nonumber\\
&-Y^d_{ij} \bar{q}_{iL} H d_{iR} -Y^u_{ij} \bar{q}_{iL} \tilde{H} u_{iR} 
-Y^e_{ij} \bar{\ell}_{iL} H e_{iR} -Y^\nu_{ij} \bar{\ell}_{iL} \tilde{H} \nu_{iR} 
\nonumber\\
&-\frac{1}{2}Y^m_{ij}\overline{\nu_{iR}^c} \phi_a \nu_{jR} \ (i,j=1,2,3)
-\frac{1}{2}Y^m_4 \overline{\nu_{4R}^c} \phi_b \nu_{4R} 
-V(\phi_a, \phi_b, H)+\mbox{H.c} .
\end{align}
where $f_i$ stand for left- and right-handed fermions, and  $H$ is the SM Higgs doublet.  $D_\mu f_i=(\partial_\mu -ig_1\tau^a
W^a_\mu -iYg_2 B_\mu -i Q_f g_F Z'_\mu)f_i$ is the covariant derivative with
$Z'_\mu$ the extra gauge boson associated with  the $U(1)_F$ gauge symmetry, 
and $g_F$ the corresponding gauge coupling constant. Since
$\phi_{a,b}$ are SM singlets, they do not play any role in the electroweak
symmetry breaking. Thus $Z'$ obtains mass only from the VEVs of the scalars
\begin{align}
  m_{Z'}^2=g_F^2 (Q_a^2 v_a^2+Q_b^2 v_b^2) .
\end{align}
From the $U(1)_F$ charge assignments in the
model, the four by four Yukawa coupling matrix is constrained to be of the block diagonal
form $\bm 3\otimes \bm 1$ in the generation space. Since the $U(1)_F$ charges
are the same for the fermions in the first three generation, there is no 
tree level FCNC induced by the $Z'$-exchange in the
physical basis after  diagonalization.  Thus  a number of constraints from 
the low energy flavor physics such as the neutral meson mixings and the $b\to s \gamma$
can be avoided.

The direct search for the process $e^+e^-\to Z'\to \ell^+\ell^-$ at the LEP-II
leads to a lower bound on the ratio of the mass to the coupling to leptons:
$M_{Z'}/g_{F} \geq 6 \mbox{ TeV}$~\cite{Carena:2004xs} for vector-like
interactions, which corresponds to a more stringent lower bound: $\sqrt{Q_a^2
  v_a^2+Q_b^2 v_b^2}\geq 6 \mbox{ TeV}$.
The current searches for narrow resonances in the Drell-Yan process
    $pp\to Z'\to \ell^+\ell^-$ at the LHC impose an alternative bound on the
    mass and the couplings of the $Z'$ boson.  
    For a model with sequential neutral gauge boson $Z'_{SSM}$ which by
  definition has the same couplings as that for the SM $Z^0$
  boson~\cite{Langacker:2008yv}, the latest lower bounds on its mass
  $M_{Z'_{SSM}}$ is $1.94$ TeV from CMS~\cite{Timciuc:2011ji} and $1.83$ TeV
  from ATLAS~\cite{Collaboration:2011dca} respectively.
  The bound on $M_{Z'_{SSM}}$ can be translated into the bound on the mass and
  couplings of the $Z'$ in this model. For instance, $g_F\lesssim 0.029$ for
  $M_{Z'}=1.44$ TeV and $g_F \lesssim 0.0051$ for $M_{Z'}=0.94$ TeV,
  respectively.

The fourth generation neutrinos obtain both Dirac and Majorana
mass terms through the vacuum expectation values (VEVs) of $H$ and $\phi_b$.  
In the basis of
$(\nu_L, \nu_R^c)^T$ the mass matrix for the fourth neutrino is given by
\begin{align}
m_\nu=\begin{pmatrix}
0 & m_D \\
m_D & m_M
\end{pmatrix} ,
\end{align}
where $m_D=Y_4^\nu v_H/\sqrt{2}$ with $v_H=246\mbox{ GeV}$ and $m_M=Y^m_4 v_{\phi_{b}} /\sqrt{2}$. 
The left-handed components  $(\nu_{1L}^{(m)}, \nu_{2L}^{(m)})$ of the two 
mass eigenstates are related to the ones in the  flavor eigenstates by a rotation angle $\theta$ 
\begin{align}
\nu^{(m)}_{1L}& =-i (c_{\theta} \nu_L -s_{\theta} \nu_R^c) , \quad
\nu^{(m)}_{2L} =s_{\theta}\nu_L +c_{\theta} \nu_R^c ,
\end{align}
where $s_{\theta}\equiv\sin\theta$ and  $c_{\theta} \equiv\cos\theta$. \note{The value of $\theta$ is defined in the range $(0,\pi/4)$} 
and  is determined by 
\begin{align}
\tan 2\theta=\frac{2m_D}{m_M},
\end{align}
with $\theta=0 \ (\pi/4)$ corresponding to the limit of minimal (maximal) mixing. 
The phase $i$ is introduced to render the two mass eigenvalues real and positive.
The two Majorana mass eigenstates are $\chi_1=\nu_{1L}^{(m)}+\nu_{1L}^{(m)c}$ and
$\chi_2=\nu_{2L}^{(m)}+\nu_{2L}^{(m)c}$, respectively.  The masses of the two neutrinos are given by
$m_{1,2}=(\sqrt{m_M^2+4 m_D^2}\mp m_M)/2$. In terms of the mixing angle $\theta$ they
can be rewritten as $m_1=(s_{\theta}/c_{\theta})m_D$ and $m_2=(c_{\theta}/s_{\theta})m_D$. 
\note{ with $m_1\leq m_2$.  Note that for all the possible values of $\theta$
  the lighter neutrino mass eigenstate $\chi_1$ consists of more
  left-handed neutrino than the right-handed one, which means that $\chi_1$
  always has sizable coupling to the SM $Z^0$ boson. Therefore the LEP-II
  bound on the mass of stable neutrino is always valid for $\chi_1$, which
  is insensitive to  the mixing angle.}

\note{ As the fermions in the first three generations and  the fourth
  generation have different $U(1)_F$ charges, the fourth generation fermions
  cannot mix with the ones in the first three generations through Yukawa
  interactions.  After the spontaneous breaking down of $U(1)_F$, there exists
  a residual $Z_2$ symmetry for the fourth generation fermions which protect
  the fourth neutrino $\chi_1$ to be a stable particle if it is lighter than
  the fourth generation charged lepton $e_4$, which makes it a possible dark
  matter candidate.}

\begin{figure}[htb]
\begin{center}
\includegraphics[width=0.65\textwidth]{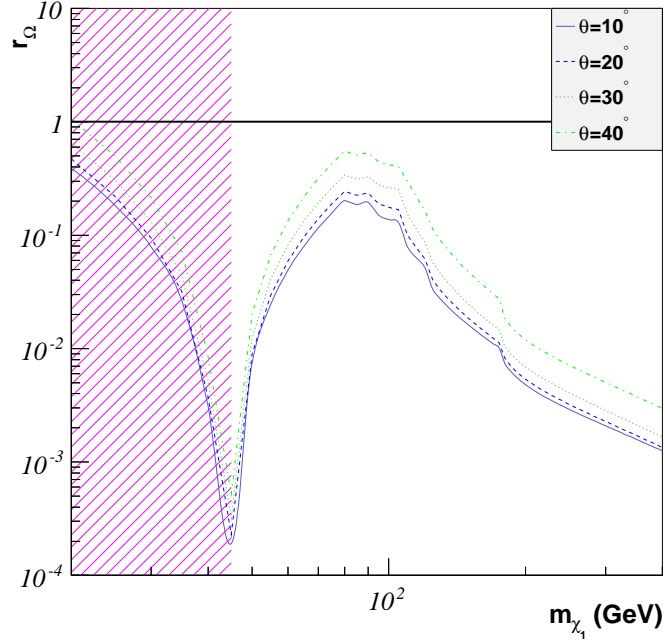}
\caption{
  The rescaled $\chi_1$ relic density  $r_\Omega$  as function of the mass of $\chi_1$.
 The shaded region is excluded by the LEP-II experiments.}
\label{fig:cross-section}
\end{center}
\end{figure}

The thermal relic density of $\chi_1$ is related to its annihilation cross section at
freeze out.  
We numerically calculate the cross sections for $\chi_1\chi_1$ annihilation
into all the relevant final states using CalHEP 2.4 \cite{Belanger:2010gh}. 
In Fig. \ref{fig:cross-section}, we show the quantity $ r_\Omega\equiv \Omega_{\chi_{1}}/ \Omega_{DM}$
 the ratio of the relic density of $\chi_1$ to the observed total DM
relic density $\Omega_{DM} h^2=0.110\pm0.006$~\cite{Nakamura:2010zzi} as
function of the mass of $\chi_1$ for different values of the mixing angle
$\theta$. The results show a significant dependence on the mixing angle
$\theta$.  For smaller mixing angle $\theta$ the couplings between $\chi_1$
and gauge bosons $W^\pm, Z$ are stronger, resulting in a smaller relic
density. The results also clearly show that due to the large annihilation cross section,
$\chi_1$ cannot make up the whole DM in the Universe. $\chi_1$ can contribute
to $\sim 20-40\%$ of the total DM relic density when its mass is around 80
GeV.  But for $m_1\gtrsim m_t$, it can contribute only a few percent or less
to the whole DM.

However, since $\chi_1$ has strong couplings to $h^0$ and $Z^0$, even in the
case that the number density of $\chi_1$ is very low in the DM halo, it is
still possible that it can be detected by its elastic scattering off nucleus
in direct detection experiments.  Given the difficulties in detecting such a
neutral and stable particle at the LHC, there is a possibility  that the
stable fourth generation neutrino could be first seen at the DM direct
detection experiments.


The generic formula for the differential event rate of DM-nucleus scattering
per nucleus mass is given by
\begin{align}\label{eq:event-rate}
\frac{dN}{dE_R}=\frac{\rho_{DM} \sigma_N}{2m_{DM} \mu_N^2}F^2(E_{R})\int^{v_{esc}}_{v_{min}}d^3 v 
\frac{f(v)}{v} ,
\end{align}
where $E_R$ is the recoil energy, $\sigma_N$ is the scattering cross section
corresponding to the zero momentum transfer, $m_{DM}$ is the mass of the DM
particle, $\mu_N=m_{DM}m_{N}/(m_{DM}+m_N)$ is the DM-nucleus reduced mass,
$F(E_{R})$ is the form factor, and $f(v)$ is the velocity distribution function
of the halo DM. The local DM density $\rho_{DM}$ is often set to be equal to 
$\rho_0\simeq 0.3 \mbox{ GeV}/\mbox{cm}^3$
which is  the local DM density
inferred from astrophysics based on a smooth halo profile. 
\note{Since the neutrino DM can only contribute to a small fraction of the
  relic density of DM, it is likely that it also contributes to a small
  fraction of the halo DM density, namely, its local density $\rho_1$ is much smaller than $\rho_0$.  
  We assume that $\rho_1$ is proportional to the
  relic density of $\chi_1$ in the Universe, namely}
\begin{align}r_\rho\equiv \frac{\rho_1}{\rho_{0}}\approx  \frac{\Omega_{\chi_{1}}}{\Omega_{DM}} ,
\end{align}
or $r_\rho\approx r_\Omega$. Consequently, the expected event rates of the DM-nucleus 
elastic scattering will be scaled down by  $r_\rho$.  In order to directly compare the theoretical
predictions with  the reported  experimental
upper limits which are often obtained under the assumption that the local DM particle density is
$\rho_0$, we shall calculate the rescaled elastic scattering cross section
\begin{align}\tilde{\sigma} \equiv   r_\rho \sigma \approx r_\Omega \sigma ,
\end{align}
which corresponds to the event rate to be seen at the direct detection experiments. 
Note that $\tilde{\sigma}$ depends on the mass of $\chi_1$ through the ratio $r_\rho$ even
when $\sigma$ is mass-independent. 
The spin-independent DM-nucleon elastic scattering
cross section in the limit of zero momentum transfer is given by~\cite{Jungman:1995df} 
\begin{align}
\sigma^{SI}_n=\frac{4 \mu_n^2}{\pi }\frac{[Z f_p+(A-Z) f_n]^2}{A^2} ,
\end{align}
where $Z$ and $A-Z$ are the number of protons and neutrons within the target
nucleus, respectively. $\mu_n=m_1 m_n/(m_1+m_n)$ is the DM-nucleon reduced mass. The couplings 
between DM and the proton (neutron) read
\begin{align}
f_{p(n)}&=\sum_{q=u,d,s}f^{p(n)}_{Tq} a_q \frac{m_{p(n)}}{m_q}
+\frac{2}{27} f^{p(n)}_{TG}\sum_{q=c,b,t}a_q \frac{m_{p(n)}}{m_q} ,
\end{align} 
with $f^{p(n)}_{Tq}$ the DM coupling to light quarks and $
f^{p(n)}_{TG}=1-\sum_{q=u,d,s}f^{p(n)}_{Tq}$. 
The coefficient $a_q$ in the model is given by
\begin{align}
a_q=c_{\theta}^2\frac{m_1 m_q}{v_H^2 m_h^2} .
\end{align}
The value of $a_q$ is proportional to $m_1$, thus larger elastic scattering
cross section is expected for heavier $\chi_1$. 
The quark mass $m_q$ in the expression of
$a_q$ cancels the one in the expression of $f_{p(n)}$. Thus there is no quark mass dependence 
in the calculations.

\begin{figure}[htb]
\begin{center}
\includegraphics[width=0.65\textwidth]{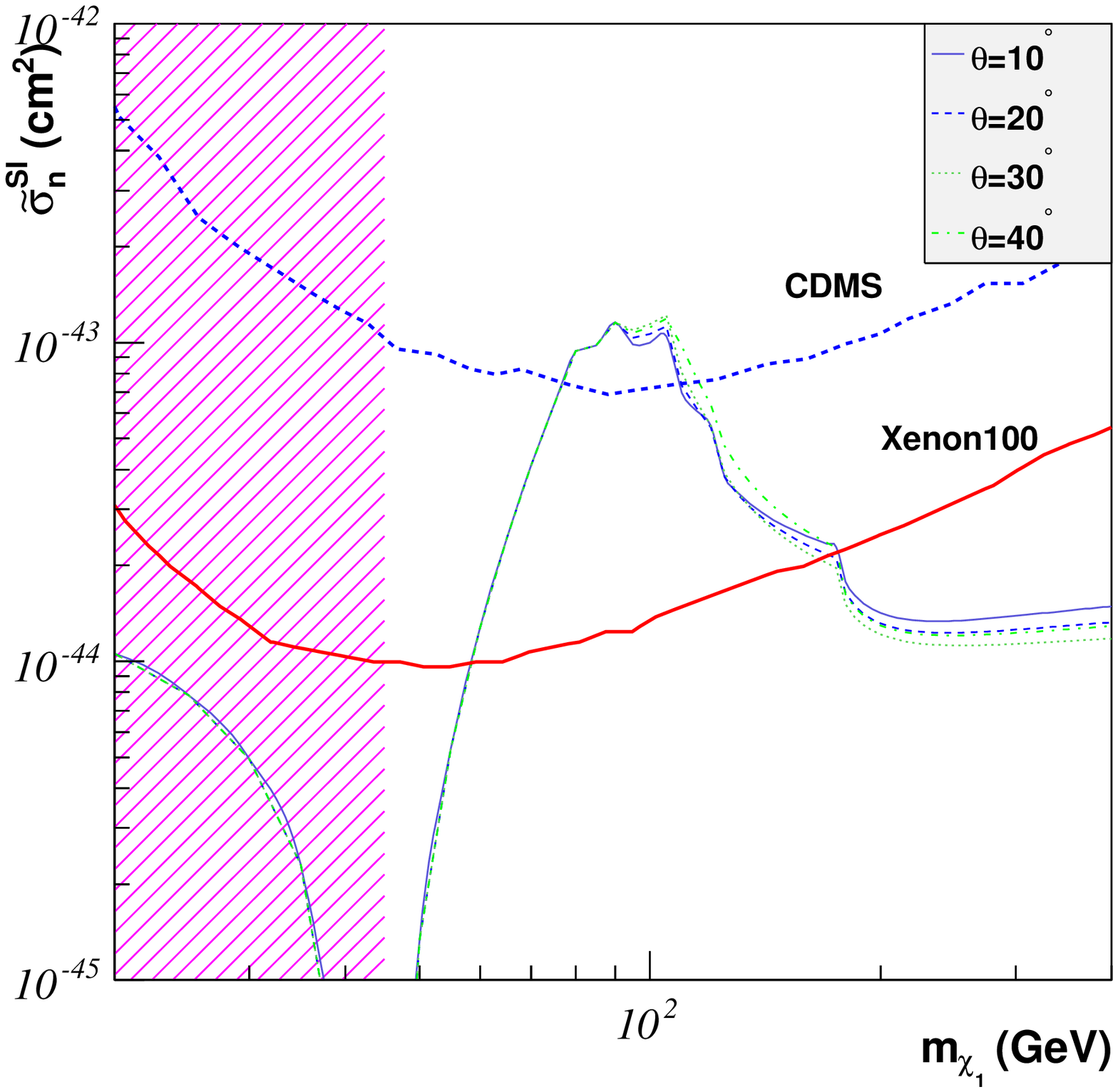}
\caption{Effective spin-independent cross section $\tilde{\sigma}^{SI}_n$
  which is $\sigma^{SI}_n$ rescaled by $r_\rho\approx r_\Omega$ for $\chi_1$
  elastically scattering off nucleon as function of the mass of $\chi_1$. Four
  curves correspond to the mixing angle $\theta=10^\circ$(solid),
  $20^\circ$(dashed), $30^\circ$(dotted) and $40^\circ$(dot-dashed)
  respectively. The current upper limits from CDMS~\cite{Ahmed:2009zw} and
  Xenon100~\cite{Aprile:2011hi} experiments are also shown.
}\label{fig:SIneutron}
\end{center}
\end{figure}

In Fig. \ref{fig:SIneutron} we give the predicted spin-independent effective cross
sections $\tilde{\sigma}^{SI}_n$ for the fourth generation neutrino 
elastic  scattering off nucleon as function of  its mass for
different values of the mixing angle $\theta$. One sees that even after the
inclusion of the rescaling  factor $r_\rho$, the current Xenon100 data can still
rule out a stable fourth generation neutrino  in the mass range 
$55\mbox{ GeV}\lesssim m_1 \lesssim 175 \mbox{ GeV}$ 
which corresponds to $r_\Omega \lesssim 1 \%$. Thus
the stable fourth generation neutrino must be heavier than the top quark,  and 
can only contribute to a small fraction of the total  DM relic density
On the other hand, for $m_{\chi_1} \gtrsim 200$ GeV, the cross
section does not decrease with $m_{\chi_1}$ increasing, and is nearly a
constant $\tilde{\sigma}^{SI}_{n}\approx 1.5\times 10^{-44}\mbox{cm}^2$ in the
range $200 \mbox{ GeV}\lesssim m_{\chi_1}\lesssim 400 \mbox{ GeV}$. This is
due to the enhanced Yukawa coupling between the fourth generation neutrino and
the Higgs boson which is proportional to $m_{\chi_1}$, as it is shown in the expression of 
$a_{q}$.  Similar conclusions are expected for other models in which DM particles interact
with SM particles through Higgs portal, for instance, the singlet scalar DM in extensions
of left-right symmetry model \cite{Guo:2011zze,Guo:2010sy,Guo:2010vy,Guo:2008si}. 
One can see from the  Fig. \ref{fig:SIneutron}  that the result is not sensitive to the mixing
angle $\theta$ either, which is due to the compensation of the similar dependencies
on $\theta$ in the relic density. \note{For instance, the cross sections for the
$W^{\pm}W^{\mp}$ and $Z^{0}Z^{0}$ channel of  $\chi\chi$ annihilation are proportional to
$c_{\theta}^4$, which compensates the $\theta$-dependence in the $a_q$ for the
elastic scattering processes.}  

The Majorana neutrino DM contributes also  to  spin-dependent elastic scattering  cross
section through  axial-vector interaction induced by the exchange of the
$Z^0$ boson. At zero momentum transfer, the spin-dependent cross section has
the following form~\cite{Jungman:1995df}
\begin{align}\sigma_N^{SD}=\frac{32}{\pi}G_F^2 \mu_n^2 \frac{J+1}{J}
\left(a_p \langle S_p\rangle + a_n \langle S_n\rangle \right)^2 ,
\end{align}
where $J$ is the spin of the nucleus, $a_{p(n)}$ is the DM effective coupling to 
proton (neutron) and $\langle S_{p(n)}\rangle$ the expectation value of the 
spin content of the nucleon within the nucleus. $G_F$ is the Fermi constant. 
The coupling $a_{p(n)}$ can be
written as
\begin{align}a_{p(n)}=\sum_{u,d,s}\frac{d_q}{\sqrt{2} G_F} \Delta^{p(n)}_q , 
\end{align}
where $d_q$ is the DM coupling to quark and $\Delta^{p(n)}_q$ is the fraction of the 
proton (neutron) spin carried by a given quark $q$. 
The coefficients $d_q$ in this model are given by
\begin{align}d_u=-d_d=-d_s=\frac{G_F}{\sqrt{2}} .
\end{align}
For the axial-vector interactions, the coupling strengths do not 
depend on the electromagnetic charges of the quarks.

\begin{figure}[htb]
\begin{center}
\includegraphics[width=0.65\textwidth]{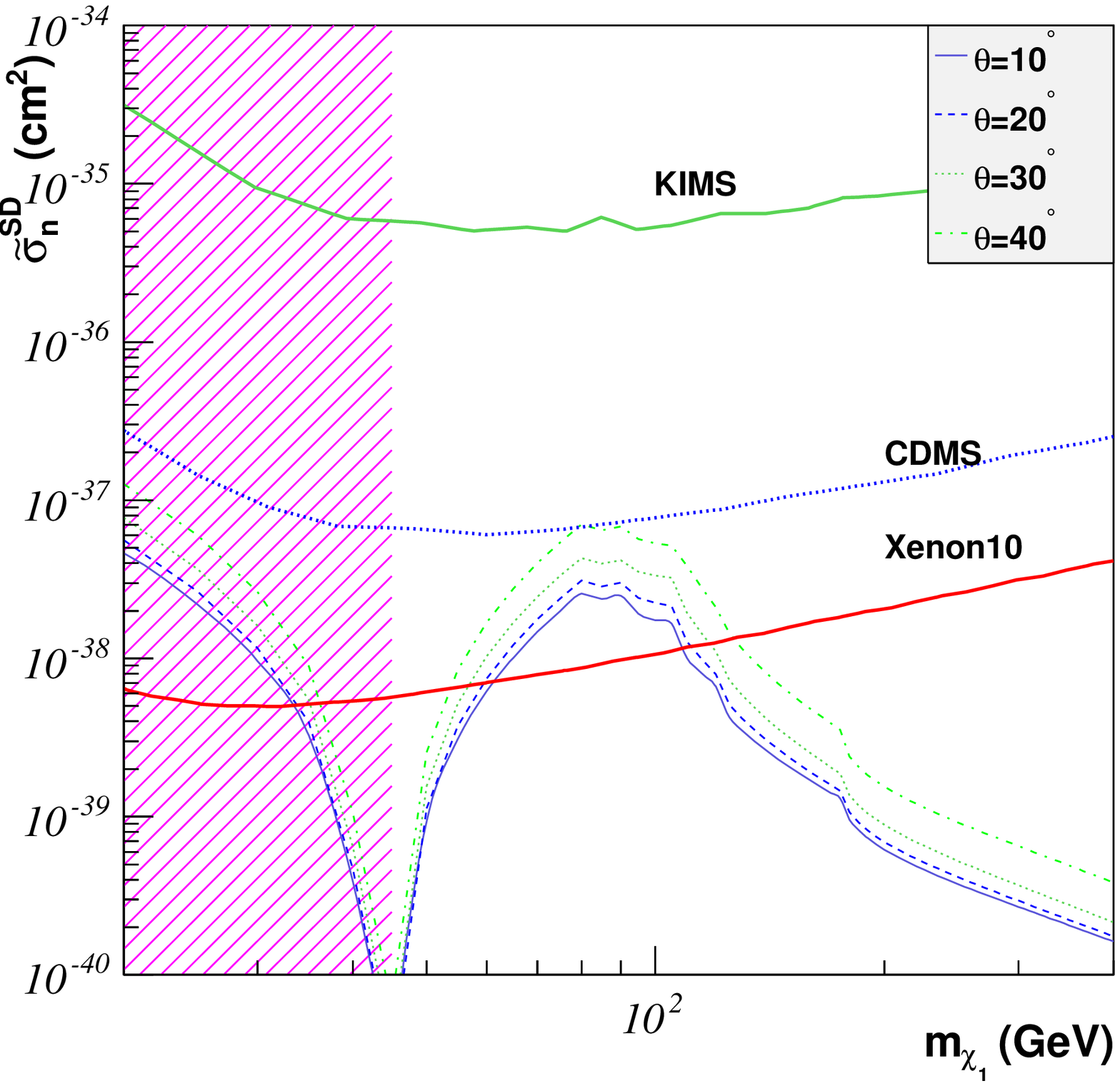}
\caption{ Effective spin-dependent cross section $\tilde{\sigma}^{SD}_n$ which
  is $\sigma^{SD}_n$ rescaled by $r_\rho\approx r_\Omega$ for $\chi_1$ elastically
  scattering off neutron as function of the mass of $\chi_1$. Four curves
  correspond to the mixing angle $\theta=10^\circ$(solid), $20^\circ$(dashed),
  $30^\circ$(dotted) and $40^\circ$(dot-dashed) respectively. The current
  upper limits from various experiments such as KIMS~\cite{Lee.:2007qn},
  CDMS~\cite{Akerib:2005za} and Xenon10~\cite{Angle:2008we} are also shown.  }
\label{fig:SDn}
\end{center}
\end{figure}

In Fig. \ref{fig:SDn} we show the predicted effective spin-dependent DM-neutron cross section
$\tilde{\sigma}_n^{SD}$ as function of the neutrino mass for different mixing
angles, together with various experimental upper limits.  Since
$\sigma^{SD}_n$ is independent of $m_{\chi_1}$, the dependency of
$\tilde{\sigma}_{p(n)}^{SD}$ on the neutrino mass comes from the dependency  of
$r_\rho$ on $m_{\chi_1}$, which can be seen by comparing Fig. \ref{fig:SDn}
with Fig. \ref{fig:cross-section}.
The Xenon10 data  is able to exclude the neutrino DM in the mass range
$60\mbox{ GeV}\lesssim m_{\chi_1} \lesssim 120\mbox{ GeV}$, which is not as
strong as that  from the Xenon100 data on spin-independent elastic
scattering cross section. For a heavy neutrino DM with mass in
the range $200\mbox{ GeV}\lesssim m_{\chi_1} \lesssim 400 \mbox{ GeV}$ the
predicted spin-dependent cross section is between $10^{-40}\mbox{ cm}^2$ and
$10^{-39}\mbox{ cm}^2$.

\begin{figure}[htb]
\begin{center}
\includegraphics[width=0.65\textwidth]{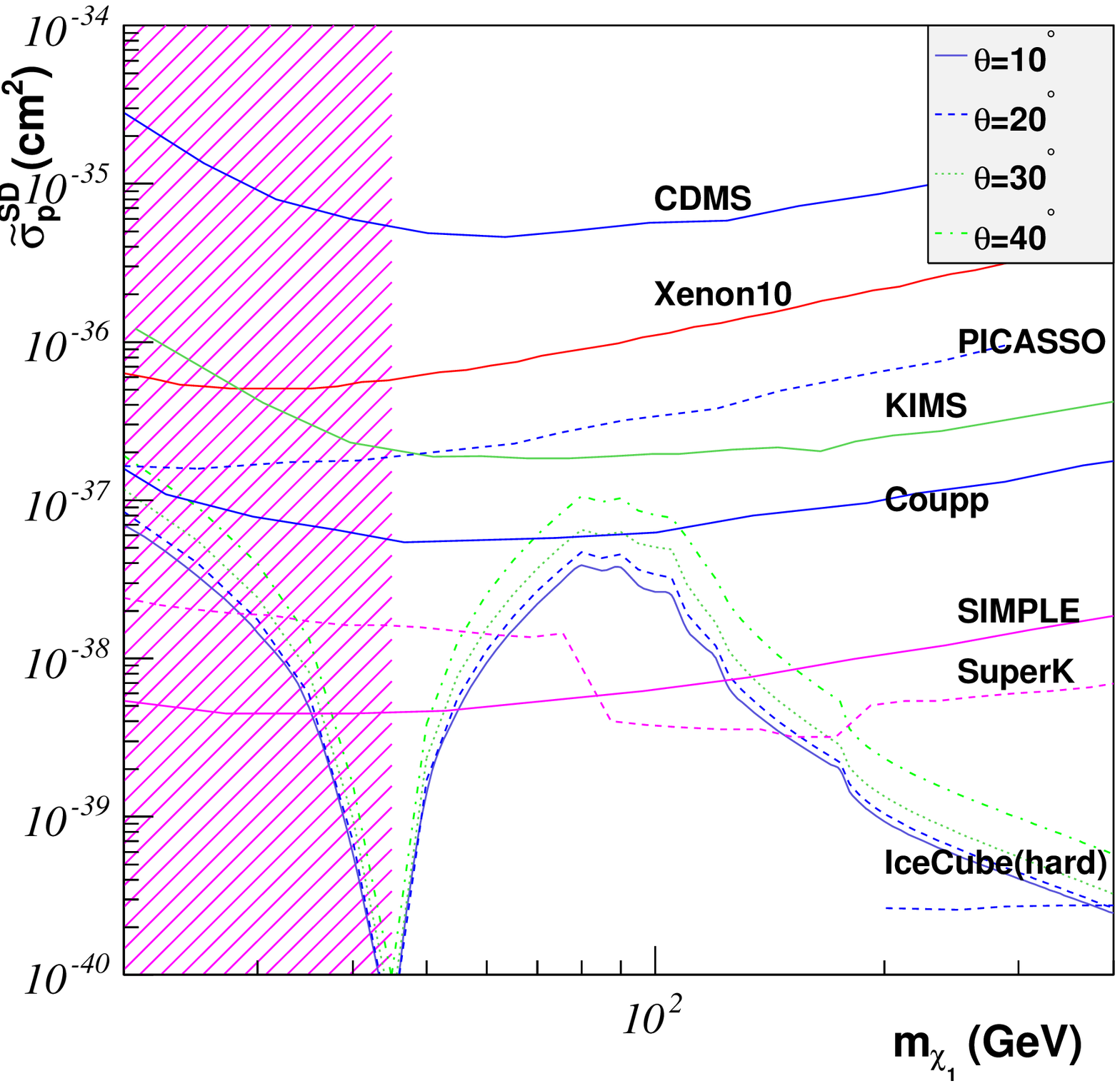}
\caption{Effective spin-dependent cross section $\tilde{\sigma}^{SD}_p$
which is $\sigma^{SD}_p$ rescaled by $r_\rho\approx r_\Omega$
  for $\chi_1$ elastically scattering off proton as function of the mass of
  $\chi_1$. Four curves correspond to the mixing angle
  $\theta=10^\circ$(solid), $20^\circ$(dashed), $30^\circ$(dotted) and
  $40^\circ$(dot-dashed) respectively.  The current upper limits from various
  experiments such as KIMS~\cite{Lee.:2007qn}, CDMS~\cite{Akerib:2005za},
  Xenon10~\cite{Angle:2008we}, Coupp~\cite{Behnke:2010xt},
  Picasso~\cite{Archambault:2009sm}, SIMPLE~\cite{Felizardo:2011uw},
  SuperK~\cite{Desai:2004pq}, and IceCube~\cite{0902.2460} are also shown.  }
\label{fig:SDp}
\end{center}
\end{figure}

In Fig. \ref{fig:SDp} we give the predicted spin-dependent DM-proton cross section
$\tilde{\sigma}_p^{SD}$. The cross sections for Majorana neutrino DM scattering off
proton and neutron are quite similar, which is due to the fact that the relative
opposite signs in $\Delta_u$ and $\Delta_d$ are compensated by the opposite
signs in $d_u$ and $d_n$. So far the most stringent limit on the DM-proton
spin-dependent cross section is reported by the SIMPLE
experiment~\cite{1106.3014}. The SIMPLE result is able to exclude the mass
range $50\mbox{ GeV}\lesssim m_{\chi_1} \lesssim 150\mbox{ GeV}$, which is compatible
with the constraints from Xenon100. 
In Fig. \ref{fig:SDp}, we also show the upper limits from indirect searches
using up-going muons which are related to the annihilation of stable fourth generation neutrinos captured in the
Sun. The limit from the Super-K experiment is obtained with the assumption
that 80$\%$ of the DM annihilation products are from $b\bar{b}$, $10\%$ from
$c\bar{c}$ and $10\%$ from $\tau\bar{\tau}$ respectively~\cite{Desai:2004pq}. In the range $170\mbox{ GeV} \lesssim m_{\chi_1} \lesssim 400\mbox{ GeV}$, the limit from
Super-K is $\sim 5\times 10^{-39}\mbox{ cm}^2$. The IceCube sets a stronger limit $\tilde{\sigma}_p^{SD} \leq 2\times 10^{-40}\mbox{ cm}^2$ for the DM mass at $250$ GeV~\cite{0902.2460}. 
This limit is obtained with the assumption that the DM annihilation products are 
dominated by $W^\pm W^\mp$. If the annihilation products are dominated by $b\bar{b}$, the 
limit is much weaker, for instance $\tilde{\sigma}_p^{SD} \leq 5\times 10^{-38}\mbox{ cm}^2$ for the DM mass at $500$ GeV~\cite{0902.2460}.  Note that in this model, the dominant final state is $Z^0h^0$. The expected limit should be 
somewhere in between. Nevertheless, the IceCube has the potential to test these predictions.


In summary, we have investigated the properties of stable fourth generation
Majorana neutrinos as dark matter particles.  Although they  contribute to a
small fraction of the whole DM in the Universe, they can still be easily probed by the
current direct detection experiments due to  their  relatively strong couplings to the SM
particles.  We have considered a fourth generation model with the stability of
the fourth Majorana neutrino protected by an additional generation-dependent
$U(1)$ gauge symmetry. 
We have shown that the current Xenon100 data constrain the mass of the stable
Majorana neutrino to be greater than the mass of the top quark. For a stable
Majorana neutrino heavier than the top quark, the effective spin-independent
cross section for the elastic scattering off nucleon is found to be
insensitive to the neutrino mass and is predicted to be around $10^{-44}
\mbox{ cm}^2$, which can be reached by the direct DM search experiments in the
near future.  The predicted effective spin-dependent cross section for the
heavy neutrino scattering off proton is in the range $10^{-40} \mbox{ cm}^2\sim
10^{-39}\mbox{ cm}^2$, which can be tested by the ongoing \note{DM
  indirect search experiments such as IceCube.}
This work is supported in part by the National Basic Research
Program of China (973 Program) under Grants No. 2010CB833000; the National
Nature Science Foundation of China (NSFC) under Grants No. 10975170,
No. 10821504 and No. 10905084; and the Project of Knowledge Innovation Program
(PKIP) of the Chinese Academy of Science.


\providecommand{\href}[2]{#2}\begingroup\raggedright\endgroup

\end{document}